\documentclass[a4paper,11pt]{article}
\usepackage[hyphens]{url}
\usepackage{hyperref}
\usepackage{graphicx}
\usepackage{subcaption}
\usepackage[capitalise]{cleveref}
\usepackage[margin=3cm]{geometry}
\usepackage[numbers,square,sort&compress]{natbib}
\usepackage{mathpazo}
\usepackage[T1]{fontenc}
\title{Containers for portable, productive and performant scientific computing}
\date{}
\author{Jack S.~Hale\thanks{Research Unit in Engineering Science,
    Faculty of Science, Technology and Communication, University of
    Luxembourg, 6 rue Richard Codenhove-Kalergi, L-1359, Luxembourg
    (\url{jack.hale@uni.lu}, \url{http://orcid.org/0000-0001-7216-861X})}
  \and
  Lizao Li\thanks{School of Mathematics, University of Minnesota,
    Minneapolis, Minnesota, 55455, United
    States. (\url{lzlarryli@gmail.com}, \url{http://orcid.org/0000-0001-7056-3368})}
  \and
  Chris N. Richardson\thanks{BP Institute, University of Cambridge,
    Madingley Road, Cambridge CB3 0EZ, United Kingdom
    (\url{cnr12@cam.ac.uk}, \url{http://orcid.org/0000-0003-3137-1392})}
  \and
  Garth N.~Wells\thanks{Department of Engineering, University of
    Cambridge, Trumpington Street, Cambridge CB2 1PZ, United Kingdom
    (\url{gnw20@cam.ac.uk}, \url{http://orcid.org/0000-0001-5291-7951})}
}

\begin{document}
\maketitle
\begin{abstract}
\noindent Containers are an emerging technology that hold promise for
improving productivity and code portability in scientific computing.
We examine Linux container technology for the distribution of a
non-trivial scientific computing software stack and its execution on a
spectrum of platforms from laptop computers through to high
performance computing (HPC) systems. We show on a workstation and a
leadership-class HPC system that when deployed appropriately there are
no performance penalties running scientific programs inside
containers. For Python code run on large parallel computers, the run
time is reduced inside a container due to faster library imports. The
software distribution approach and data that we present will help
developers and users decide on whether container technology is
appropriate for them. We also provide guidance for the vendors of HPC
systems that rely on proprietary libraries for performance on what
they can do to make containers work seamlessly and without performance
penalty.
\end{abstract}
\section{Introduction}
\label{sec:introduction}

Traditionally, scientific and engineering software libraries had few
dependencies and were developed by small groups in a stable and
well-established language, such as~Fortran~77 or~C. In contrast, many
modern scientific libraries have numerous dependencies and are
developed by multiple groups and developers in various (and
increasingly often multiple) languages, e.g.~C++, C, Fortran, Python,
R.  In modern scientific software libraries, key tasks are commonly
delegated to specialised, high-performance libraries, and these form
the dependencies of higher-level libraries.  These lower-level
libraries will in turn themselves typically have dependencies.  The
introduction of dependencies, which usually differ in their level of
maintenance, standards compliance and build system quality, brings
significant challenges for users installing scientific libraries, and
for portability of scientific libraries to different platforms.

Introducing high-quality, specialised software dependencies into a
library can allow small research teams to produce novel, sophisticated
and high performance programs with fewer errors and reduced
development time, and without being expert in the implementation of
all underlying operations.  However, effective distribution of complex
software packages and their dependencies, both within a research group
and to the wider community, is challenging and time consuming. It is
difficult to justify the researcher time required to support the
deployment of software on esoteric, wide-ranging and sometimes poorly
maintained platforms, from laptops to high performance computing
facilities, through to cloud computing infrastructures. Moreover, the
necessary expertise and range of test systems is not generally
available in research teams for actively supporting a wide range of
platforms.

Motivated by the difficulties in distributing modern high performance
software to a wide range of users and developers, we address the use
of Linux containers as an approach to portable deployment of
scientific software libraries, from laptop computer to supercomputer,
with no noticeable performance penalty. We describe a container-based
approach that we have developed for both users and developers of a
collection of widely used scientific libraries from the FEniCS Project
(\url{https://fenicsproject.org}), an open-source finite element
method (FEM) toolkit for solving partial differential equations. We
use a common image for the FEniCS libraries in tutorials for new users
(running on their own computers) through to high performance, parallel
simulations.  The focus of this work is the ease and simplicity with
which containers can be used to dependably share and deploy scientific
software stacks, without performance penalty, across a range of
platforms.  We want to share our experiences to aid the developers of
other scientific libraries in deciding whether to adopt containers to
support the distribution of their software.

To demonstrate the performance characteristics of container-based
approaches for scientific applications, we present a range of
performance data on different systems comparing native and
containerised performance. In particular, we address running on HPC
systems that require system-specific vendor libraries for performance.
We also touch upon what steps vendors of HPC systems, which often
depend on proprietary libraries for performance (most notably Message
Passing Interface (MPI) libraries), can take to support the deployment
of container technologies on HPC systems.

Containers have already received some attention in the academic
literature.  \citet{boettiger:2015} focused on scientific
reproducibility by using containers as a stable framework to propagate
software stacks between generations and teams of researchers.
\citet{apaolaza:2016} used Docker~\citep{docker:www} to package and
execute a complex video processing workflow, with an accompanying blog
post by \citet{haines:2016} containing specifics on using Zenodo
(\url{https://zenodo.org}) to archive software with a permanent
Digital Object Identifier (DOI).  \citet{tommaso:2015} and
\citet{gerlach_skyport:2014} examined the use of containers to ease
the portability challenge of complex data analysis workflows, where
multiple containers (each containing its own complex software
environment) are chained together. \citet{tommaso:2015} also compared
the performance of some commonly used genetics analysis software
running natively and inside a container.  \citet{seo:2014} compared
the performance of containers and virtual machines for non-scientific
software stacks deployed in the cloud.  \citet{macdonnell:2007}
benchmarked the performance of the VMWare virtualisation platform for
a variety of common scientific computing tasks. The overhead for
compute-intensive tasks was around 6\% and for IO tasks around
9\%. \citet{felter:2015} also looked at the performance differential
of non-scientific software within virtualised and containerised
environments. The performance studies published in the literature
\citep{felter:2015, tommaso:2015, seo:2014} found negligible
differences between container and native performance, as long as the
container is setup and used appropriately, e.g.~using data mounts for
I/O intensive tasks \citep{felter:2015} and not setting up many
containers for individual tasks with run times on the order of
container setup time (typically fractions of a
second)~\citep{tommaso:2015}.

The focus of this paper differs from previous studies in two ways.
Firstly, instead of concentrating on how containers aid scientific
reproducibility~\citep{apaolaza:2016, haines:2016, boettiger:2015}, we
consider how containers can be used to ease the difficulty of sharing
and distributing scientific code to developers and end-users on a wide
range of computing platforms. Secondly, unlike previous studies
\citep{tommaso:2015, seo:2014, felter:2015} that showed performance
parity between containerised and native \emph{shared} memory parallel
software, we consider performance of software using \emph{distributed}
memory parallel programming models (e.g.~MPI). MPI forms the backbone
of modern HPC systems and modern scientific computing codes in the
physical sciences, so for containers to be accepted by the community
the performance characteristics on distributed memory machines must be
demonstrated.

\section{Background: Linux containers}
\label{sec:background}

We give a brief overview of containers by comparison with traditional
virtual-machine based virtualisation, a technology which many
researchers in the scientific computing community are already familiar
with, an we provide a simple example of configuring a container.  Our
objective is not to provide an in-depth review of the technology
behind Linux containers; we refer the reader to the
white-paper~\citep{docker-security:2015} for a detailed discussion of
a container runtime and security model.

\subsection{Technology}

Containers are, on the surface at least, similar to traditional
virtualisation technologies, e.g.~VMWare, QEMU, Hyper-V, Xen, in that
they allow multiple isolated applications to run on a single host
computer. Traditional virtualisation involves running an entire guest
operating system (OS) within an isolated environment, with the kernel
of the guest OS communicating with the host via a complete set of
virtualised devices (CPU, disk, memory, etc.). In contrast, Linux
containers share the underlying kernel of the host and isolate the
processes running within it from all other processes running on the
underlying host operating system. Instead of virtualising the
hardware, containers virtualise the operating system
itself~\citep{merkel:2014}.

While the end effect of both traditional virtualisation and containers
are similar, namely, an isolated, distinct and portable computing
environment, the means by which they achieve this results in a very
different experience for the end-user. Traditional virtual machines
load an entire guest operating system into memory. Each guest OS might
take up gigabytes of storage space on the host, and require a
significant fraction of system resources just to run, even before
doing any work. Furthermore, starting up each guest OS usually takes
on the order of minutes.  Because containers share the existing host
kernel, they can utilise most of the dependencies and features of the
host. If a container does not execute anything, it uses little to no
system resources.  Running thousands of containers on one host is
possible. To start a new container, the host just has to start new
processes that are suitably isolated from its own. This means that
startup times are similar to that of a native application.

We wish to note the distinction between an image and a container.  An
image is an immutable file that contains a description of a complete
computing environment (libraries, executables, default networking
configuration, default initialisation, and more) that can be used by a
runtime (e.g.,~Docker) to create a container. A container is a runtime
instantiation of a particular image. It is possible to have many
containers based on the same image.  Each new container can re-use the
data stored in an image without duplication, just storing the
differences in a layered file system.  We want to emphasise that the
image format and container runtime are mutually exclusive
entities. Currently, by far the most popular container runtime is
Docker~\cite{docker:www}. However, we show performance results in this
paper using the Docker, rkt~\citep{rkt:www} and
Shifter~\citep{shifter:www} runtimes, each of which uses a different
approach to instantiate containers from the same image. The recently
formed Open Container Initiative~\citep{oci:www} seeks to standardise
image formats and container runtimes across vendors so that images and
containers are portable between the different runtime platforms.

In summary, containers can initially be viewed as a lightweight
counterpart to traditional virtualisation technology. Traditional
virtualisation still has some advantages over containers, such as a
time-proven security model.  Containers, by virtue of sharing the host
OS kernel directly, cannot run images designed for another OS. So if a
user wishes to run Linux images in a container on macOS, then they
must still use a virtual machine to host an OS with a Linux kernel.

\subsection{Elementary demonstration}

To provide an simple exposition of the use of images and containers
with the Docker runtime, we show how to write a {\tt Dockerfile},
build an image from it, and finally how to launch a container from the
created image.

Simply put, a {\tt Dockerfile} file is a sequence of some
Docker-specific directives and shell script commands that prescribe
the entire build process and runtime configuration of a software
environment. An attractive feature of {\tt Dockerfile} files is that
they unambiguously describe the configuration and build process behind
an image.

An example of a simple {\tt Dockerfile} that builds an Ubuntu
16.04-based image with with SciPy installed from the system package
manager might be:
\begin{verbatim}
FROM ubuntu:16.04
USER root
RUN apt-get -y update && \
    apt-get -y upgrade && \
    apt-get -y install python-scipy && \
    rm -rf /var/lib/apt/lists/* /tmp/* /var/tmp/*
\end{verbatim}
Running {\tt docker build .} in the same directory as the {\tt
  Dockerfile} will build and package the result into an image with a
unique mathematical hash, e.g.~{\tt d366b85a7fdb}.  The image can be
given an easy to remember name using {\tt docker tag d366b85a7fdb
  scipy-image}. Using
\begin{verbatim}
docker run -ti scipy-image python
\end{verbatim}
an interactive Python terminal session will appear and the SciPy
module will be available.  The above methodology can be extended to
package and distribute complex scientific software stacks with the
benefit of working with a well-defined environment.  We have only
shown three {\tt Dockerfile} directives ({\tt FROM}, {\tt USER} and
{\tt RUN}) in the above example, many more are discussed in the
Docker documentation at \url{https://docker.com}.

A strength of both the Open Container Initiative and Docker image
formats is the layered file system. Taking the above {\tt Dockerfile}
as an example, the directive {\tt FROM ubuntu:16.04} instructs the
build to start with the official Ubuntu Docker image as the first
layer. This first layer is often called a base image. Many other base
images are available in registries, for example at Dockerhub
(\url{https://hub.docker.com}) and Quay (\url{https://quay.io}). Every
subsequent directive (here, {\tt USER} and {\tt RUN}) creates a new
layer that only stores the difference between itself and the previous
layer.  Each layer is associated with a unique mathematical hash.
This simple but elegant construct has multiple beneficial effects.
First, if we build stages of a software pipeline in separate images
with each {\tt FROM} the same base image (e.g.~{\tt ubuntu:16.04}),
the end-user only needs to download the base image once and storage of
the complete pipeline can be very compact.  Secondly, on calling {\tt
docker run} to run the image in a new container, a new empty
filesystem layer is created to store runtime changes made by the user.
This operation is very lightweight, taking up only a few kilobytes of
disk space in addition to the storage space required by the
modification, and typically takes fractions of a second to run.

\section{Distributing, sharing and reproducibility with scientific software}

We now discuss some issues that open-source scientific software
projects face when distributing code and supporting user installation,
and how container runtimes and standardised image formats can help
address these issues. Much of what we present is influenced by our
first-hand experiences as developers of the open-source FEniCS Project
(\citep{fenics:book}, \url{http://fenicsproject.org}).

\subsection{Installation and complex software dependency trees}

Users of scientific software run on a broad range of platforms,
including Windows, macOS and a variety of Linux distributions, with
their installations maintained to varying degrees.  Moreover, users
have different needs; from beginners wishing to run their first
analysis, to advanced users wanting a specific version of a library or
wanting to run on high-performance computing clusters, through to
developers adding new features.  The complex chain of dependencies
that characterises many modern, user-level scientific libraries
compounds the difficulties in supporting robust and appropriately
configured installations. Supporting installation on all user
platforms and meeting the needs of all users is beyond the resources
of the vast majority of open-source scientific software projects.

Up until early 2015 the FEniCS Project supported an automated
installation system called `Dorsal'.  Now deprecated, Dorsal was a
somewhat successful attempt to build an entire FEniCS environment
automatically on a variety of Unix-like operating systems. Notably
absent was Windows support. As much as possible, Dorsal used binary
software dependencies from distributions (packages), and then compiled
(and where necessary patched) various other dependencies and FEniCS
itself from source on the end-user's machine.  In practice, this
approach has a number of issues.  There are a wide variety of Linux
distributions and each is subtly different, which in practice means
supporting even a small number of distributions reliably is a
significant task requiring time and a range of test
systems. Furthermore, even small customisations by a user to their
underlying environment, like having a different version of a library
or application in their {\tt PATH}, could lead to a failure of the
build.

Container runtimes, such as Docker, rkt and Shifter, largely abstract
away the differences between the many operating systems that users
run. The same image runs identically inside a container on almost any
platform.  Using container images, we now
distribute an entire FEniCS environment that has proven to be
reliable, and is simple and quick to run on wide variety of systems.
The environment is reproducible, in that every image is associated
with a mathematical hash that is unique to every specific realisation
of a build. This has greatly simplified development and testing
processes, and improved user experience.  FEniCS users of the images
have required very little support, and when support has been required
is has become easier to provide as the environment in which a user is
working is well-defined.

Containers complement, rather than replace, packaging and build
systems such as apt, pip and EasyBuild. We make heavy use of both apt
and pip when building the FEniCS images, with only judicious use of
builds directly from source. The advantage of using these existing
systems within a container environment is that the build system is
completely isolated from the user's underlying system. In our
experience, this leads to significantly fewer build failures. In
addition, developers can choose the best build systems for the job at
hand, rather than being limited by what system administrators have
installed or what end-users are comfortable with.

\subsection{Ease of launching a runtime environment}
\label{sec:ease}

We want every new user of FEniCS to be able to reliably run it on
almost any platform with a minimal number of commands.  Once users
have installed Docker on their platforms, they can start using FEniCS
with the command:
\begin{verbatim}
docker run -ti quay.io/fenicsproject/stable
\end{verbatim}
To share files from the host into the container environment the user
can add the {\tt -v}~flag:
\begin{verbatim}
docker run -ti -v $(pwd):/home/fenics/shared quay.io/fenicsproject/stable
\end{verbatim}

While the above commands are straightforward, many users and
developers will want more control over the environment,
e.g.~persistent state and the ability to re-launch a container.
However, the command line interface (CLI) to the Docker runtime has
been designed with technical users in mind. Moving beyond the simplest
interactions, the CLI is too low-level for typical scientific software
users and developers, and especially for beginners.  For example,
launching an interactive Jupyter notebook~\citep{PER-GRA:2007} session
requires knowledge of the {\tt docker} command's networking features:
\begin{verbatim}
docker run -w /home/fenics/shared -v $(pwd):/home/fenics/shared -d \
  -p 127.0.0.1:8888:8888 quay.io/fenicsproject/stable \
  `jupyter-notebook --ip=0.0.0.0'
\end{verbatim}

To make the use of the containers as easy as possible for users
following typical workflows we have developed a lightweight wrapper
script {\tt fenicsproject} that sets up some common workflows, such as
starting a Jupyter notebook session as described above:
\begin{verbatim}
fenicsproject notebook my-project
fenicsproject start my-project
\end{verbatim}
This script has been used successfully for a number of FEniCS
tutorials, with participants using the FEniCS Python interface via
Jupyter notebooks on their own laptop computers. FEniCS developers are
also using this script to quickly setup and build development branches
of FEniCS in lightweight isolated environments.

For users who want to develop their own FEniCS container workflows
using the {\tt docker} command we provide extensive
documentation~\citep{containerdocs} that targets common scientific
computing use cases.

\subsection{Deployment on high performance computing systems}

One of the most appealing features of containers is the possibility of
running the same image on a wide range of computing platforms,
including HPC systems.  However, a recurring challenge when moving to
a new HPC machine is configuring and compiling a library with all its
dependencies. A characteristic of HPC systems is that each is unique
and the software environment changes over time as compilers and
libraries are updated. Moreover, users and developers will typically
work infrequently on HPC systems, with development and testing
typically taking place on user machines before occasional execution of
large-scale simulations on HPC systems. It is our experience that
there are serious productivity barriers when moving from local
development machines to HPC systems. We find that, particularly with
infrequent HPC use, there is a high time cost in updating and
rebuilding libraries, even for experienced developers.  Due to the
differences between HPC systems, there is also a significant time cost
associated with moving from one HPC system to another.  For
experienced users, these issues lower productivity; the time to update
a software stack on a HPC system can dwarf the wall clock run time.
For less experienced users, these issues can seriously hamper
accessibility. The appeal of containers is that a user's familiar
local development environment can be reproduced with minimal effort
and time on a HPC system.

Shifter~\citep{nersc:2015} is a new open-source project at the
National Energy Research Scientific Computing Centre (NERSC) that
provides a runtime for container images, and is specifically suited to
the security and performance challenges of HPC workflows.  The Docker
image format is one of a number of image formats supported by Shifter.
Images are pulled ahead of running a job, by issuing the {\tt
  shifterimg pull} command, which pulls an image from an image
registry. The practical task of running a container using Shifter is
then straightforward.  For example:
\begin{verbatim}
shifter --image=docker:quay.io/fenicsproject/stable:2016.1.0r1 \
./demo_poisson
\end{verbatim}
will run the command {\tt ./demo\_poisson} inside a FEniCS container.
In contrast to the Docker runtime, the Shifter images are run in
read-only mode and the user's environment and home directory are
automatically passed through to the container. When using the Docker
runtime, users can store data inside a container using the layered
file system, whereas with Shifter any user generated objects,
e.g.~executable files, must be stored outside of the container on the
host system.  We refer to \citet{nersc:2015} for further details.

When executing a program in parallel, MPI is typically used to
communicate (pass messages) between processes. The MPI libraries
included in the FEniCS image work adequately if a parallel job is run
on a single compute node because modern MPI libraries do not use the
network to pass messages when memory is shared.  However, when a job
is distributed across more than one compute node, communication takes
place across a network and the standard MPI libraries in the container
will not take advantage of the specialised high-performance networking
hardware that characterises modern HPC systems.  Thanks to the MPICH
application binary interface (ABI) initiative~\citep{mpich:2013}, we
can distribute an image linked against the Ubuntu MPICH library
package, and change the linking of the libraries to the host
ABI-compatible MPI implementation at runtime. We discuss this in more
detail in the run time performance section.

\subsection{Provisioning and maintenance of images}
\label{sec:daytoday}

To build the entire FEniCS software stack we maintain a collection of
{\tt Dockerfile} files under version control at
\url{https://bitbucket.org/fenics-project/docker}.  If a user wants to
install their own pieces of software alongside FEniCS, they can create
their own simple {\tt Dockerfile} that begins with {\tt FROM
  quay.io/fenicsproject/stable}. We also provision the images with
scripts for users to update and build specific libraries within a
container.

\begin{figure}
  \center
  \includegraphics[width=0.8\textwidth]{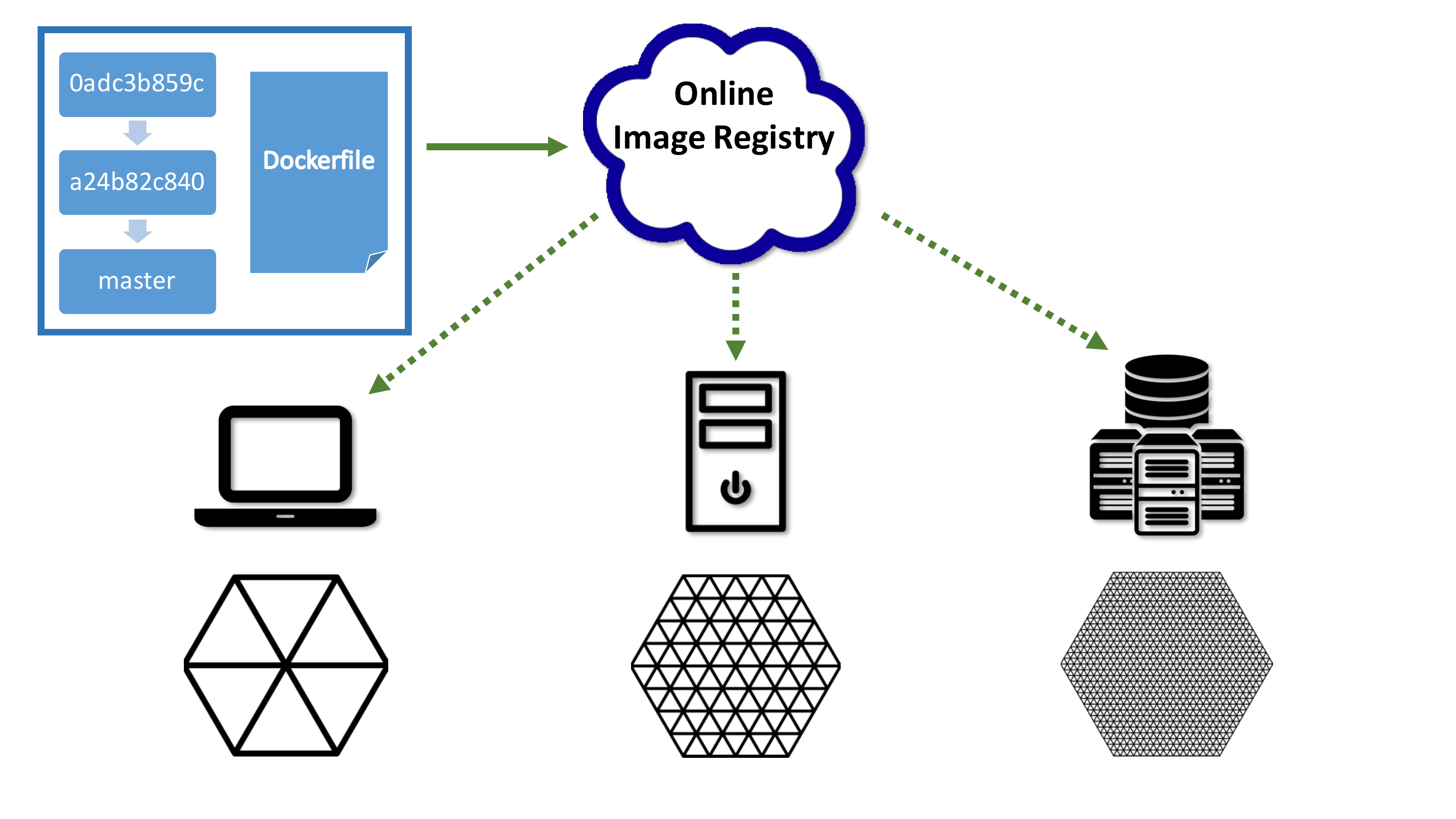}
  \caption{An image is defined by a Dockerfile, and is built locally
    or by a cloud-based service such as \url{https://quay.io}. The
    latter can also also act as a registry. An image is pulled from
    the registry and run on different platforms, for example solving
    smaller problems on a laptop, to large-scale problems on an HPC
    machine.}
  \label{fig:images}
\end{figure}

We use a cloud-based service (\url{https://quay.io}) to build and
distribute FEniCS images. We have developed a hierarchy of images that
allow re-use between various end-user containers, e.g.~stable user,
debug and developer images.  On a commit to the repository where we
store the {\tt Dockerfile} files, a build command is executed
automatically. When the build is finished, the images, in both Docker
and Open Container Initiative-compatible formats, are ready and
available publicly. We have found this workflow very reliable and
requiring very little developer intervention.  Making small
configuration changes requires changing just one file and the steps
used to build the container are automatic and transparent.  The
process from Dockerfile to deployment of an image on different systems
is illustrated in \cref{fig:images}.

\section{Runtime performance studies}
\label{sec:performance}

Performance is a critical factor in deciding whether or not containers
are viable for scientific software. We address this point by
presenting performance data for a range of representative simulations
on a typical workstation, and tests and benchmarks on a leadership
class HPC system. We pay special attention to use of system MPI
libraries on the HPC system.

All necessary instructions, scripts and images, along with the raw
data and plotting scripts from our own tests, are available in the
supporting material~\citep{supporting:www}.

\subsection{FEniCS workstation performance}
\label{sec:performance-docker}

We compare the performance of a selection of tests running inside the
FEniCS Project stable image, hosted at
\texttt{quay.io/fenicsproject/stable:2016.1.0r1}, running within the
Docker~\citep{docker:www} and rkt~\citep{rkt:www} runtimes to a nearly
identical version of FEniCS running on the same machine natively.  In
addition, we also run the same image with the Docker runtime inside a
virtual machine~(VM), which reflects the experience of macOS and
Windows users where a lightweight Linux distribution must be
virtualised to host Docker.  The Docker virtual machine setup on macOS
and Windows is transparent to the user and their experience is nearly
identical to that of Docker users on Linux.

As far as possible, we create identical software stacks within the
image and natively on the host. We use the Ubuntu 16.04 LTS Linux
distribution as the base in both cases, and use as many core libraries
from the distribution packages (e.g.,~MPICH, OpenBLAS, LAPACK) as
possible.  The test system is a 16-core Intel Xeon-based workstation
(2 $\times$ E5-2670 (Sandy Bridge)) with 128GB of RAM. The relative
differences in performance between each of the environments are of
interest.

Within each environment we run a selection of single-process finite
element method performance tests, that is, \emph{without} using MPI.
We briefly summarise these tests here: `Poisson LU' solves a
two-dimensional Poisson equation using LU decomposition, `Poisson AMG'
solves a three-dimensional Poisson equation using the conjugate
gradient method preconditioned by algebraic multigrid, `IO' reads in a
large mesh file from the host and writes a solution back to a file on
the host, and finally `elasticity' solves a three-dimensional
elasticity problem using the conjugate gradient method with an
algebraic multigrid preconditioner.  These tests are slightly modified
versions of demos distributed with the FEniCS Project and are
representative of the types of simulations that users might run.  The
test programs are implemented using the Python interface of FEniCS.
FEniCS uses just-in-time (JIT) compilation from Python to generate
loadable shared objects at runtime for operations that are
model-specific, and these shared object libraries are cached as files.
The run times reported below do not include the JIT compilation time,
which is only incurred on the first run.

\begin{figure}
  \center
  \includegraphics[width=0.95\textwidth]{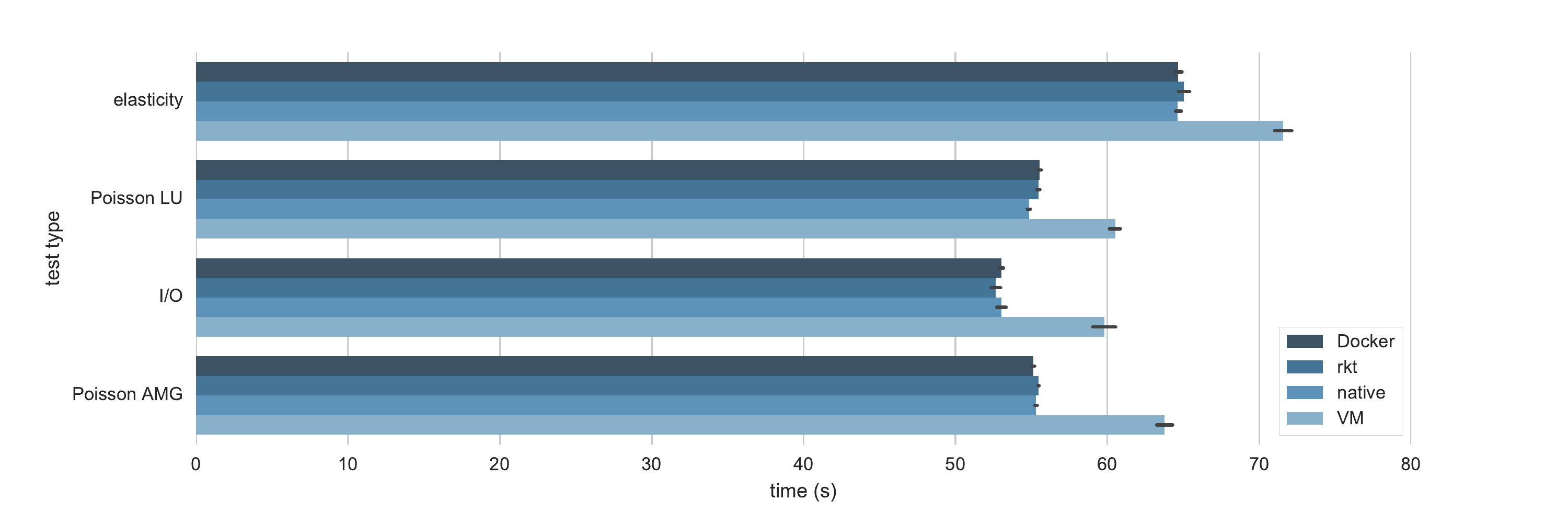}
  \caption{Performance of \emph{Docker}, \emph{rkt}, \emph{native} and
    \emph{VM} platforms running four different benchmark problems on a
    workstation. Shorter bars are better. \emph{Docker}, \emph{rkt}
    and \emph{native} are very close, with performance differentials
    less than~1\%. The extra overhead introduced by the \emph{VM}
    virtualisation leads to a performance drop of around~15\%.  The
    error bars for five runs are shown in black.}
  \label{fig:end-user}
\end{figure}

\Cref{fig:end-user} shows the run times for the four tests in the four
environments. The \emph{Docker}, \emph{rkt} and \emph{native}
platforms provide nearly identical performance.  This is no surprise
given that the container runtimes leverage the same Linux kernel
functionality. However, it does highlight that there are a variety of
runtime systems to choose from, and choices can be made to suit
particular system policies.  The \emph{VM} platform has up to a 15\%
performance penalty, slightly worse that the results of
\citet{macdonnell:2007}. Other virtualisation environments may provide
better results than the VirtualBox software we used in this work. We
conclude that for these tests that there is no performance penalty
when running inside a container.

\subsection{FEniCS HPC performance tests using the Shifter runtime}
\label{sec:performance-shifter}

FEniCS supports distributed memory parallelism (using MPI), and has
been shown to scale to over 20,000 processes on HPC
machines~\citep{figshare:2015}.  We have prepared a simple FEniCS test
program which solves the Poisson equation using the conjugate gradient
method with an algebraic multigrid preconditioner, and which also
involves distributed mesh refinement and I/O. We use this to compare
native and containerised performance.  The test program is fairly
representative of a typical user program. This test program is
implemented in both the C++ and Python interfaces of FEniCS.

The tests in this section have been performed on \emph{Edison}, a Cray
XC30 system at NERSC. Edison has 24 cores per compute node (2 $\times$
E5-2695v2 (Ivy Bridge)). To use the system MPI libraries, we need the
containerised application to load the Cray MPI libraries. We do this,
as shown by Bahls~\citep{bahls:cug2016}, by adjusting the {\tt
  LD\_LIBRARY\_PATH} environment variable on job submission and
copying the required system libraries to a path that is accessible
from within the container.  This hinges on ABI compatibility of the
container and system MPI libraries. The command
\begin{verbatim}
srun -n 192 shifter env LD_LIBRARY_PATH=$SCRATCH/hpc-mpich/lib \
  --image=docker:quay.io/fenicsproject/stable:2016.1.0r1 ./demo_poisson
\end{verbatim}
uses the SLURM submission system MPI launcher to launch Shifter on 192
cores and adds the appropriate library path to the environment. The
required libraries for using the system MPI are copied into {\tt
  \$SCRATCH/hpc-mpich/lib}, a location that is visible from within a
running container. All other libraries are provided by the container,
and are identical to those that we distribute to end-users for running
on laptops and workstations.  Note that with Shifter we execute the
MPI {\tt srun} command on the host and each process is a container.

To test performance we compare a native build of the FEniCS library
and test application to a container build. The native build uses the
Cray system modules as much as possible (gcc/4.9.3, cray-mpich/7.2.5,
libsci/16.07.1, cray-tpsl/16.03.1, cray-petsc/3.6.1.0). The only
FEniCS dependency compiled from source was SWIG, which is required
when using the Python interface.  For the C++ version of the container
test, the test code was compiled inside the container on Edison by
invoking an interactive Shifter session. For the Python version of the
test, the JIT cache of shared objects was pre-generated and the
reported run times do not include the JIT compilation time.

\begin{figure}
  \center
  \includegraphics[width=0.95\textwidth]{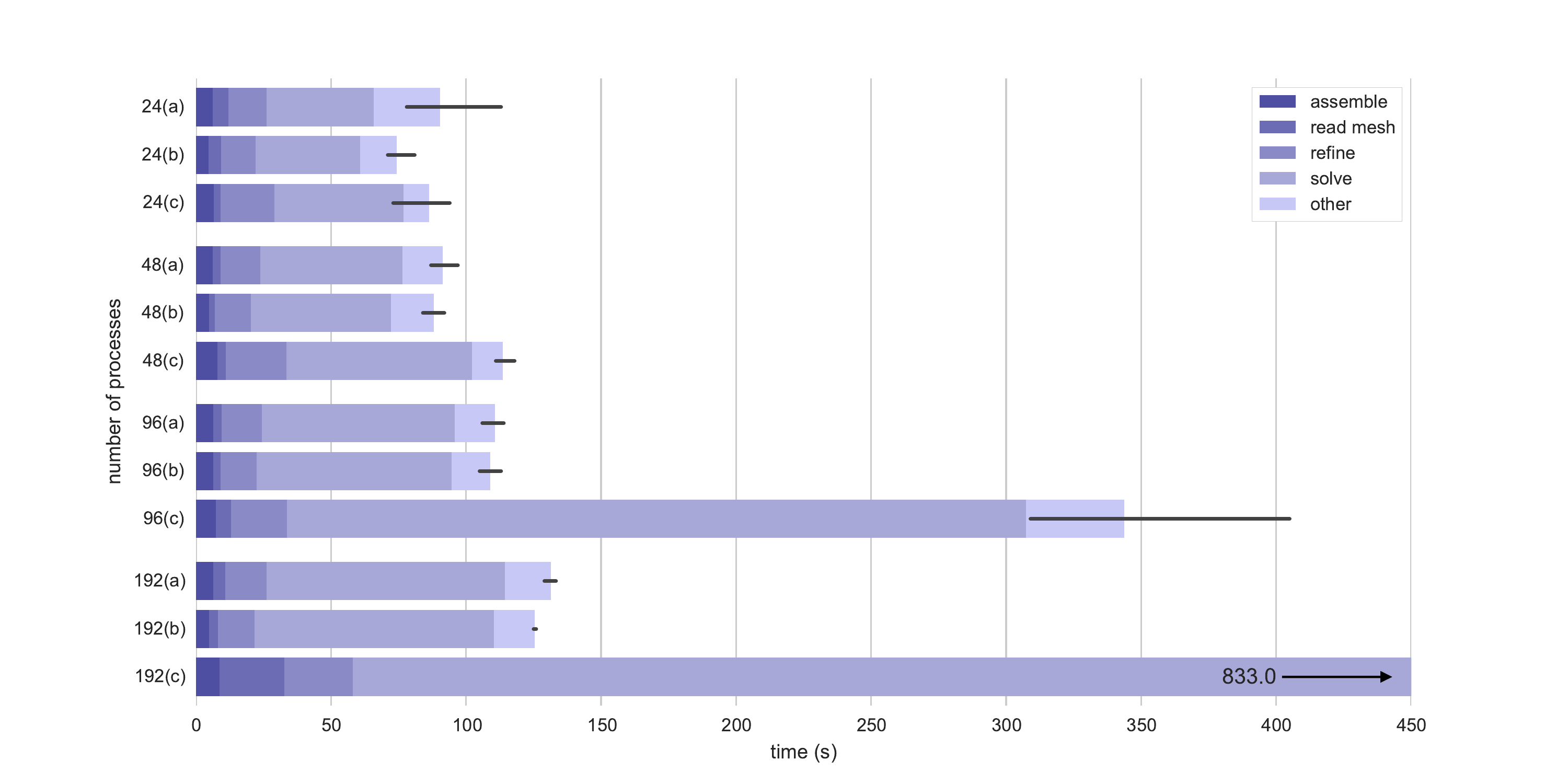}
  \caption{C++ FEniCS benchmark run times on \emph{Edison} when:
    (a)~run natively without containers; (b)~run in Shifter with Cray
    MPI libraries; and (c)~run in Shifter with the container MPI
    libraries. Tests were performed with with 24, 48, 96 and 192 MPI
    processes, with one MPI process per CPU core. Shorter bars are
    better.  The $x$-axis is truncated for the 192(c) result.  In each
    case, the time is shown for various compute operations. There is
    little difference between the run times for cases (a) and~(b),
    which both use the system MPI library.  Performance for (c)
    deteriorates rapidly when computing across compute nodes (Edison
    has 24 cores per compute node). Error bars for three runs are
    shown in black.}
  \label{fig:cpp-shifter}
\end{figure}

Wall clock run times for the C++ version of the test program on Edison
are shown in \cref{fig:cpp-shifter} for between 24 and 192 processes.
Performance inside the container when using the system MPI
library~(case (b) in \cref{fig:cpp-shifter}) is comparable to or
better than the native performance~(case (a) in
\cref{fig:cpp-shifter}). When using the container MPICH library (case
(c) in \cref{fig:cpp-shifter}), the solver time increases dramatically
for higher MPI process counts, highlighting the importance of loading
the system MPI library for container programs.

Python has become increasingly popular in HPC, in particular as a
`driver' for programs written in compiled languages. There is however
a serious performance problem on HPC systems, which is often referred
to as the `Python import problem', see~\citep{pythonimport:2014}.  The
Python {\tt import} statement appears early in a Python program, and
each process imports (loads) Python modules/code from disk into
memory. Each imported module will itself typically import/load modules
from disk. HPC filesystems are optimised for access to a few large
files by many processes, but can be extremely slow when many processes
access many small files, as happens when Python imports modules on
each process. When running with 1000 processes we have observed on
some systems that it can take over 30 minutes to import the Python
modules required by the Python interface of FEniCS.  Containers bypass
this problem as the Python modules are included in the filesystem of
the container image, which is mounted on each node as a single large
file~\citep{nersc:2015}. A small price is paid in memory usage per
node.

\begin{figure}
  \center
  \includegraphics[width=0.95\textwidth]{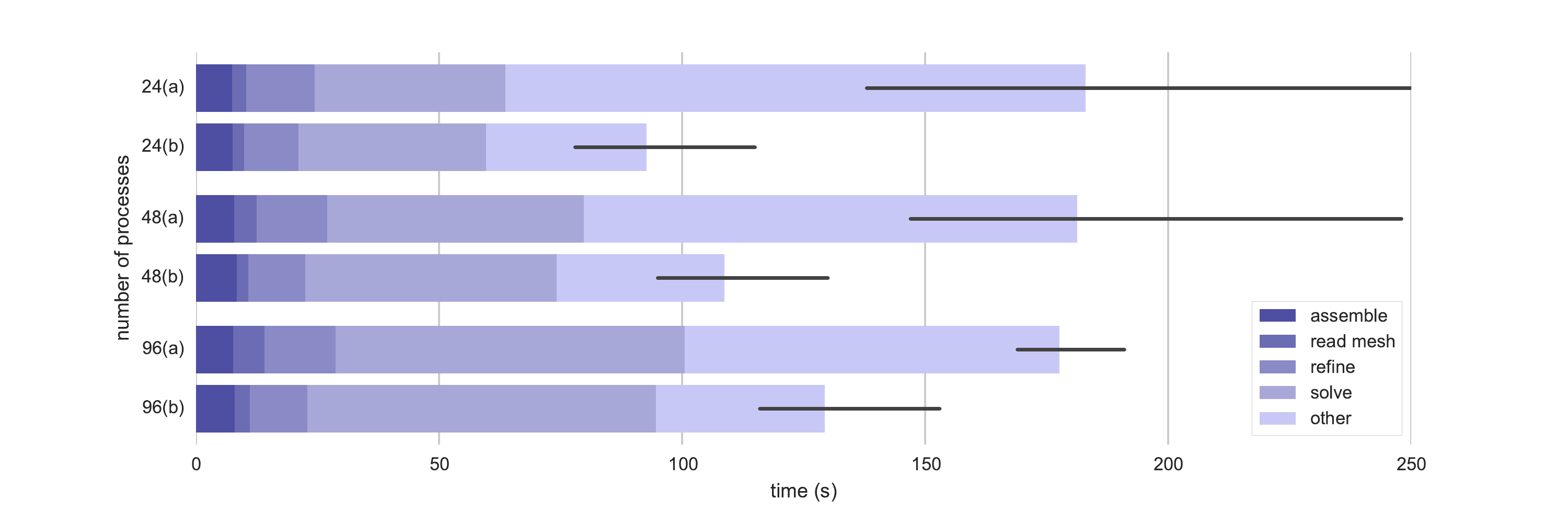}
  \caption{Python benchmark run times on \emph{Edison} when: (a) run natively
    without containers; and (b) run inside Shifter with the system MPI
    library. Tests were performed with with 24, 48 and 96 MPI
    processes, with one MPI process per CPU core. Shorter bars are
    better. Each compute operation takes comparable time in each case,
    but the total run time is significantly greater and more variable
    when running natively because of the Python module import
    time. Error bars for three runs shown in black.}
  \label{fig:python-shifter}
\end{figure}

To examine the improvement in Python import performance,
\cref{fig:python-shifter} shows the run times for native and
containerised (using the system MPI library) executions of the Python
version of the test program.  The key observation here is the large
difference in total run time between the native (case (a)) and
container (case (b)) executions of the program, with run time for the
native case being significantly greater than the container case. This
is due to the Python import time. Additionally, the native run time
shows more variation, which is probably due to unpredictable
contention on the file system.  We also note that containers have an
advantage for Python programs that employ just-in-time compilation
since we can be sure that the necessary compilers and libraries are
available on the compute nodes.

\subsection{High-Performance Geometric Multigrid (HPGMG-FE) benchmark}
\label{sec:hpgmg}

HPGMG-FE is a new benchmark for ranking
supercomputers~\citep{HPGMGv1}. It aims to provide a balanced measure
of system performance characteristics that are important for modern
scientific software on HPC systems. It is a useful test for container
performance due its highly tuned implementation; any weakness in
configuration or container performance issues become acute and
immediately obvious. For optimal performance, the benchmark requires
exploitation of a CPU-specific instruction set, namely the Advanced
Vector Extensions (AVX).

In the same way as our C++ benchmarks above, we have compiled HPGMG-FE
both natively and in a container. In both cases, it is important to
compile for the host architecture and to use the best optimisation
flags available for the C compiler. Thus we ensure that features of
the host CPU, such as the AVX instruction set, are activated.  We make
use of OpenBLAS, which dynamically selects at runtime the
implementation that is appropriate for the CPU type.  For workstation
tests we execute the MPI {\tt mpirun} command inside a container,
whereas on Edison the MPI {\tt srun} command is executed from outside
the containers.

\begin{figure}
  \center
  \begin{subfigure}[b]{\textwidth}
    \center \includegraphics[width=0.95\textwidth]{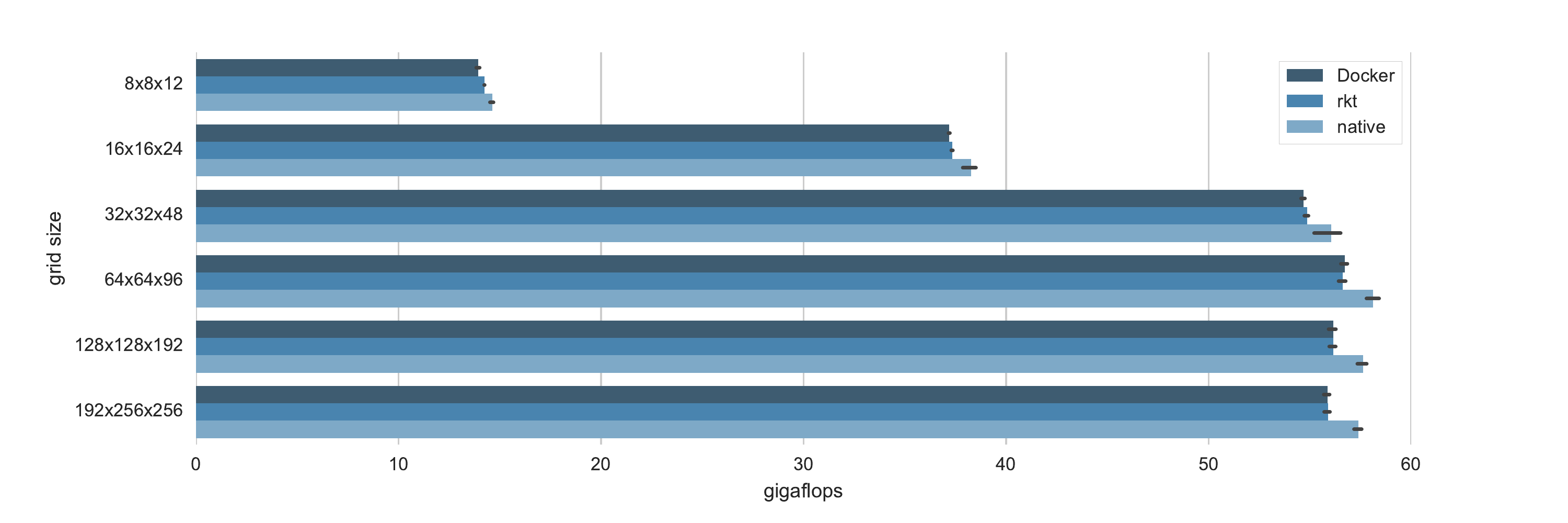}
    \caption{A 16-core Xeon workstation with the \emph{Docker},
      \emph{rkt} and \emph{native} platforms.}\label{fig:hpgmg-galah}
  \end{subfigure}
  \begin{subfigure}[b]{\textwidth}
  \center
  \includegraphics[width=0.95\textwidth]{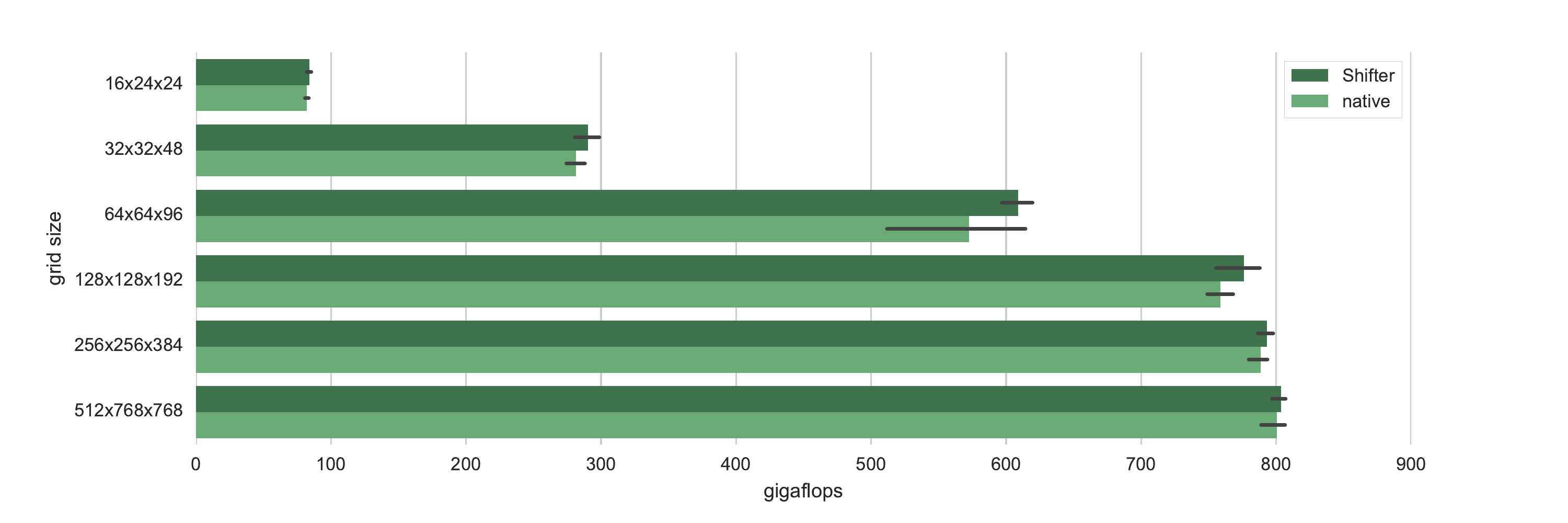}
  \caption{Edison Cray-XC30 using 192 cores with the \emph{native} and
    \emph{Shifter} platforms.}\label{fig:hpgmg-edison}
  \end{subfigure}
  \caption{HPGMG-FE performance on (a) Intel Xeon and (b) Cray XC30.
    Longer bars are better. This difference between the native and
    containerised performance is small. Error bars for five runs.}
  \label{fig:hpgmg}
\end{figure}

The performance results for the HPGMG-FE benchmark are shown in
\cref{fig:hpgmg} using 16 cores on the Xeon workstation
(\cref{fig:hpgmg-galah}) and using 192 cores on Edison
(\cref{fig:hpgmg-edison}). On the workstation, the native compile
outperforms the Docker and rkt containers by around 3\%, while on
Edison the Shifter container essentially matches the native
performance at larger problem sizes.  Both performance margins are in
our view small enough to claim that the containers match the native
performance of both systems.

This benchmark is an example of where a precompiled program or library
inside a container might not be able to exploit hardware instructions
that are specific to the host architecture and which may be critical
for performance (vectorisation in this case). In such cases a
container can be provisioned with the necessary scripts to build
performance critical binaries on the host system. Such scripts are
straightforward to develop and support as they are run inside the
well-defined and controlled environment of the container.

\section{Conclusions}

We have discussed some of the challenges in distributing scientific
software and the role that containers can fulfil in addressing this
issue. Images can provide a complete, stable, and consistent
environment that can be easily distributed to end users, thereby
avoiding the difficulties that end-users commonly face with deep
dependency trees.  Containers have the further advantage of largely
abstracting away the host system, making it possible to deliver a
common and consistent environment on many different platforms, be it
laptop, workstation, cloud instance or supercomputer. We have shown
through a selection of test problems (both representative and
benchmark) using a common image that containers introduce no
performance penalty; on the contrary it is our anecdotal experience
that optimised images generally outperform native user installations
as the container environment can be suitably optimised by the
developers of a library.  A production or development environment can
be moved with ease and efficiency between platforms. This will improve
productivity in the use of different systems for scientific computing.

We have shown performance data including tests on a Cray XC30 HPC
system. The test data underlines the importance of using the native
MPI library on HPC platforms to fully exploit the specialised
interconnect.  Loading the Cray MPI library from a container was made
straightforward by the ABI compatibility between the Cray MPI library
and the open-source MPICH library.  We hope that our results will
strengthen arguments for vendors to make their proprietary and
specialised system libraries ABI compatible with open-source
libraries, and to devise simple approaches to switching dynamically
linked libraries. This will enhance the usability of HPC systems via
containers without compromising performance.  We also showed that
containers provide a portable solution on HPC systems to the problem
with languages that load a large number of small files on each process
(the `Python import problem'). Images also provide a robust solution
to just-in-time compilation on HPC systems by encapsulating the
necessary libraries and compilers, which are sometimes not available
on HPC compute nodes.

We hope that the discussion and performance data in this work will
assist researchers in scientific computing who are contemplating an
investment in container technology. We believe that container
technology can improve productivity and sharing in the scientific
computing community, and in particular can dramatically improve the
accessibility and usability of HPC systems.

\subsubsection*{Supplementary material}

Supporting instructions, data, images and scripts are archived
at~\citep{supporting:www}.

\subsubsection*{Acknowledgements}

We thank Linh Nguyen (Melior Innovations) and the National Energy
Research Scientific Computing Center (NERSC) for their assistance in
using Edison, and Don Bahls at Cray for advice on using Shifter with
MPI.  JSH is supported by the National Research Fund, Luxembourg, and
cofunded under the Marie Curie Actions of the European Commission
(FP7-COFUND) Grant No.~6693582.  LL is supported in part by NSF grant
DMS-1418805.  CNR is supported by the Engineering and Physical
Sciences Research Council (UK) through grant agreement EP/N018871/1.
\bibliographystyle{abbrvnat}
\bibliography{references}
\end{document}